\documentclass[10pt,oneside,reqno]{amsart}
\usepackage{amsfonts,amssymb,
amscd,amsmath,latexsym,amsbsy,bm}
\numberwithin{equation}{section}

 \usepackage[usenames,dvipsnames]{xcolor}

\usepackage[colorlinks,hyperindex, 
            bookmarks=true,bookmarksopen=true]{hyperref}
    \hypersetup
    {
        colorlinks=true,
        linkcolor=MidnightBlue,
        urlcolor=MidnightBlue,
        filecolor=black,
        citecolor=RedOrange,
        pdfstartview=FitV,
        pdftitle={},
        pdfauthor={Ilmar Gahramanov, Kemal Tezgin},
        pdfsubject={},
        pdfkeywords={},
        pdfpagemode=None,
        bookmarksopen=true
    }

\usepackage{mathtext}
\usepackage{stackrel}

\usepackage[utf8]{inputenc}

\begin{document}

$\;$\\ 
\vspace{2.7cm}

\centerline{\LARGE \bf A resurgence analysis for cubic and quartic anharmonic potentials}
\vskip 1.3 cm

\centerline{\large {\bf Ilmar Gahramanov$^{a,b,c}$ and Kemal Tezgin$^{a, d}$}  }

{\small
\begin{center}
\textit{$^a$Max Planck Institute for Gravitational Physics (Albert Einstein Institute),\\ Am M\"{u}hlenberg 1, D-14476 Potsdam, Germany} \\
\vspace{2.5mm}
\textit{ $^b$Institute of Radiation Problems ANAS,\\ B.Vahabzade 9, AZ1143 Baku, Azerbaijan} \\
\vspace{2.5mm}
\textit{$^c$Department of Mathematics, Khazar University, \\ Mehseti St. 41, AZ1096, Baku, Azerbaijan} \\
\vspace{2.5mm}
\textit{$^d$Department of Physics, University of Connecticut, \\ Storrs, CT 06269, USA} \\
\texttt{} \\
\vspace{.1mm}
\vspace{.1mm}
\href{mailto:ilmar@aei.mpg.de}{{\textbf{ilmar.gahramanov@aei.mpg.de}}} $\;\;\;\;\;\;\;\;$  \href{mailto:kemal.tezgin@uconn.edu}{{\textbf{kemal.tezgin@uconn.edu}}}
\end{center}
}

\vskip 0.5cm 

\centerline{\bf Abstract} \vskip 0.2cm \noindent In this work we explicitly show resurgence relations between perturbative and one instanton sectors of the resonance energy levels for cubic and quartic anharmonic potentials in one-dimensional quantum mechanics. Both systems satisfy the Dunne-\"{U}nsal relation and hence we are able to derive one-instanton non-perturbative contributions with the fluctuation terms to the energy merely from the perturbative data. We confirm our results with previous results obtained in the literature. \\ \vspace{0.4cm}


\newpage

\section{Introduction}

Perturbation theory, even often very useful, misses many interesting physical phenomena like exponentially suppressed instanton effects. In usual perturbative expansion, the energy is expanded as
\begin{equation}
    E(g) = \sum_{n=0}^\infty E_n g^n \;,
\end{equation}
where $g$ is the coupling constant\footnote{In quantum mechanics the semi-classical expansion is in powers of the Planck constant.}. As the coupling constant goes to zero, terms like $e^{-\frac{S}{g}}$ cannot be represented in this expansion for any finite positive real number $S$. Moreover, in many interesting physics problems the perturbation theory is mathematically ill-defined: In large orders, the coefficients $E_n$ grow like a factorial of the order and not alternating by sign which leads to the fact that the expansion is not Borel summable. Hence the resummation procedure leads to ambiguous imaginary terms. In fact, these missing physical effects and the mathematically ill-defined picture of the series are related to each other. Once we include these non-perturbative effects to the perturbative expansion in a trans-series form, the ambiguous imaginary terms cancel each other and the mathematically ill-defined picture is removed. Typically once we involve the instanton effects we get the following form of trans-series expansion for the $N$th energy level
\begin{equation} \label{fullexpansion}
E^{(N)}(g)=\sum_{n=0}^\infty \sum_{l=1}^{n-1}\sum_{m=0}^\infty c_{n, l, m}\, \frac{e^{-n\frac{S}{g}}}{g^{n(N+\frac{1}{2})}}\hskip -.1 cm  \left(\ln \left[\frac{a}{g}\right]\right)^l \hskip -.1 cm g^{m}\; ,
\end{equation}
with coefficients $c_{n, l, m}$ and constant $a$. This expansion has three generators: the expansion parameter $g$, a single instanton factor $e^{-\frac{S}{g}}$ with instanton action coefficient $S$ and logarithmic terms arising from the instanton interaction. Note that $n=0$ corresponds to the usual perturbative expansion and the logarithmic terms first start appearing at two-instanton sector.  

In order to derive the full expansion of the form (\ref{fullexpansion}), in a series of works \cite{ZinnJustin:2004ib, ZinnJustin:2004cg, Jentschura:2010zza, ZinnJustin:1989mi} Zinn-Justin et al. have conjectured generalized quantization conditions whose small-g expansion leads to the full trans-series expansion. In simple cases these generalized quantization conditions depend on two functions $B(E,g)$ and $A(E,g)$. The function $B$ is called the perturbative function since it is obtained purely from perturbative expansion and its solution for $E(B,g)$ reduces to the usual Rayleigh-Schr\"{o}dinger perturbative expansion of the $N$th energy level once the value of $B$ is restricted to $N+\frac{1}{2}$. The function $A(E,g)$ is called the non-perturbative instanton function which includes the instanton action as the first term plus the instanton  fluctuation terms. Interested reader can find more details about these functions in \cite{ZinnJustin:1989mi}.

For many systems with degenerate minima including double-well and sine-Gordon potentials, Dunne and \"{U}nsal have shown by using the uniform WKB approach \cite{Dunne:2013ada,Dunne:2014bca,Dunne:2016qix} (see also \cite{Misumi:2015dua} for the sine-Gordon potential) that the following relation holds between $A(E,g)$ and $B(E,g)$ functions 
\begin{equation}
\frac{\partial E(B, g)}{\partial B} =  -\frac{g}{2S} \left(2B+ g \frac{\partial A(B,g)}{\partial g} \right) \;,
\end{equation}
where the energy $E$ and the instanton function $A$ are expressed in terms of $B$ and $g$. This relation has also been shown to hold for the resonance energy levels of a particle located at the local minima of cubic and quartic anharmonic potentials \cite{Gahramanov:2015yxk}. In such systems the particle has a non-vanishing decay width and the instantons are associated with the imaginary part of the energy.

One may easily recognize that the Dunne-\"{U}nsal relation has a powerful consequence: Rather than calculating both perturbative and non-perturbative sectors separately, it is enough actually to compute one sector in order to derive the other. Hence from the generalized quantization condition, it is actually enough to know one of the functions $A, B$ in order to derive the full-resurgent expansion. Therefore this relation implies a high correlation between the coefficients $c_{n, l, m}$ and has an explanation in the framework of resurgence analysis. 

In this work we summarize a few results obtained recently in \cite{Gahramanov:2015yxk}, i.e. we review the relation between perturbative and non-perturbative sectors for cubic and quartic anharmonic potentials. We discuss the implications of this relation by explicitly deriving one-instanton contributions including the fluctuation terms to the energy levels for the mentioned potentials by using only the perturbative calculations. Our results confirm the known results in the literature. We restrict the discussion to the cubic and quartic anharmonic oscillators only and follow the notations used by Zinn-Justin et al. in the papers \cite{ Jentschura:2010zza, Jentschura:2009fok}. 

There is a wide literature on cubic and quartic anharmonic potentials. Here we discuss the energy levels of these potentials from the resurgence point of view. After completing this work, it came into authors' attention\footnote{We thank Marcos Marino for pointing out these works to us.} that there is an overlap between this work and the references \cite{Alv1,Alv2,Alv3}. 

\section{Generalized quantization conditions}

In this section we give a brief review of the perturbative WKB method developed in \cite{ZinnJustin:2004ib, ZinnJustin:2004cg, Jentschura:2010zza, ZinnJustin:1989mi}. In these works Zinn-Justin et al. proposed a  complete description of the energy eigenvalues including the non-perturbative effects by using a generalized quantization condition which is an extension of the usual Bohr-Sommerfeld quantization condition. An equal description of Bohr-Sommerfeld quantization condition can be obtained as follows. Let us consider time-independent Schr\"{o}dinger equation for cubic anharmonic potential
\begin{equation}
    \left(-\frac12 \frac{\partial^2}{\partial q^2}+V(q) \right) \phi = E \phi \:,
\end{equation}
with $V(q)  = \frac12 q^2 +\sqrt{g} q^3$. Once we apply the scaling $q\rightarrow g^{-\frac12}q$ and define a function $s(q)=-g\frac{\phi'(q)}{\phi(q)}$, we are able to write the Schr\"{o}dinger equation in the form of the Riccati equation as follows
\begin{equation}\label{Riccati}
    gs'(q)-s^2(q)+q^2+2q^3-2gE=0 \;.
\end{equation}
Here by fixing $E$ and $g\rightarrow 0$ allows us to define the perturbative expansion of $s(q)$ whose coefficients are recursively related \cite{Jentschura:2010zza}. One important property of the function $s(q)$ is on the complex plane
\begin{equation}\label{Bohr}
    -\frac{1}{2\pi ig}\oint dz s(z)=\frac{1}{2\pi i}\oint dz\frac{\phi'(z)}{\phi(z)}=N \;,
\end{equation}
where $N$ is the energy level. The equation ($\ref{Bohr}$) is the Bohr-Sommerfeld quantization condition and directly follows from the number of zeroes that a wave function has. Since the $N$th state harmonic oscillator wave function has $N$ zero values, a small perturbed anharmonic oscillator has the same number of zeroes. Furthermore, by decomposing $s(q)$ into antisymmetric and symmetric parts $s(q) = s_{odd}+s_{even}$ under simultaneous transformations $g\rightarrow -g$ and $E\rightarrow -E$, one finds that the antisymmetric part $s_{odd}$ can be purely expressed by the symmetric part as follows
\begin{equation}
    s_{odd} = \frac{g}{2} \frac{s_{even}'(q)}{s_{even}(q)} \;.
\end{equation}
Then we can write the Bohr-Sommerfeld quantization condition ($\ref{Bohr}$) equivalently in the following way
\begin{equation}\label{Bohr-Sommerfeld}
-\frac{1}{2 \pi i g} \oint dq s_{even}(q) = N+\frac12 \;.
\end{equation}
From equation ($\ref{Riccati}$), it is clear that the function $s_{even}$ is implicitly dependent on $g$ and $E$. Hence we define the function $B_3(E,g)$ for cubic potential as
\begin{equation}
B_3(E,g)=-\frac{1}{2 \pi i g} \oint dq s_{even}(q).
\end{equation}
A similar function $B_4(E,g)$ for quartic potential can be defined by the same procedure. To include the instanton effects to the energy, we need a different quantization condition than ($\ref{Bohr-Sommerfeld}$). The generalized Bohr-Sommerfeld quantization condition, we are going to mention next, is developed for that reason. In order to get the instanton contributions, in \cite{Jentschura:2010zza} an additional function $A(E,g)$ is needed for cubic and quartic anharmonic  potentials. In their approach, the generalized quantization conditions read
\begin{equation} \label{cubicquant}
\frac{1}{\Gamma\left( \frac12 - B_3(E,g)\right)} =
\frac{1}{\sqrt{8 \pi}}\, \left( - \frac{8}{g} \right)^{B_3(E,g)} \,
e^{-A_3(E,g)} \, ,
\end{equation}
\begin{equation}
\frac{1}{\Gamma\left( \frac12 - B_4(E,g)\right)} =
\frac{1}{\sqrt{2 \pi}} \left( \frac{4}{g} \right)^{B_4(E,g)} \; e^{-A_4(E,g)} \; ,
\end{equation}
for cubic and quartic potentials, with $g>0$ and $g<0$ respectively. The functions $B$ and $A$ have the following expanded forms
\begin{align} \label{Bfor}
B(E,g) & = E+\sum_{i=1}^\infty g^i b_{i+1}(E) \;,\\ \label{Afor}
A(E,g) & = S_{\text{instanton}}+\sum_{i=1}^\infty g^i a_{i+1}(E) \;,
\end{align}
where $b_i$ and $a_i$ are polynomials of degree $i$ in $E$ and $S_{\text{instanton}}$ denotes the instanton action. In (\ref{Bfor}) solving $E$ for $B=N+1/2$ will return us the usual perturbative expansion for the $N$th energy level. The $A$ function (\ref{Afor}) consists of single instanton action and instanton fluctuation terms. In fact, the calculation of the function $A$ is in general more challenging than the perturbative function $B$. Since our aim is to derive the one-instanton fluctuation terms by using the $B$ function only, we will not discuss the derivation of the $A$ function and refer interested reader to \cite{ZinnJustin:1989mi}. Instead, we will recover the fluctuation terms by using the Dunne-\"{U}nsal relation.

\section{Fluctuation factors}    

In this section we present the procedure to compute the fluctuation factors of one instanton sector for energy eigenvalues of cubic and quartic anharmonic oscillators.

\subsection{Cubic anharmonic oscillator}
The Hamiltonian we consider for the cubic anharmonic osciallator is
\begin{equation} \label{cubicH}
\mathbb{H}(g) = -\frac{1}{2} \frac{\partial^2}{\partial q^2}  + \frac{1}{2}  q^2 + \sqrt{g} q^3. 
\end{equation}
Here we consider the case for $g>0$. Then the system possess resonances and the resonance energies are complex\footnote{In case $g<0$ the Hamiltonian (\ref{cubicH}) is $PT$-symmetric and has a real spectrum, see, e.g. \cite{ZJim}.} where the complex part of the energy is associated with instanton configurations \cite{Jentschura:2010zza}. A single instanton action for this system is $\frac{2}{15g}$. \\

The usual Rayleigh-Schr\"{o}dinger perturbative expansion of the ground state energy for this potential yields the following expansion \cite{Vainshtein, Drummond}
\begin{equation}{\label{pertcubic}}
E_{\rm ground}(g) = \frac12 -\frac{11}{8} g -\frac{465}{32} g^2 - \frac{39709}{128} g^3 -
\frac{19250805}{2048} g^4 + \ldots \, .
\end{equation}
The coefficients of this expansion is non-alternating for $g>0$ and in large orders the coefficients grow like a factorial of the order. Hence the series is non-Borel summable.

Once we include the instanton effects to the system, in \cite{Jentschura:2010zza, JZcubic} Zinn-Justin et al. discussed the expansion of the quantization condition (\ref{cubicquant}) leads to the following resurgent expansion of the $N$th energy level \cite{JZcubic,Jentschura:2010zza}
\begin{equation} \label{cubicexpansion}
E^{(N)}(g)=\sum_{n=0}^\infty E^{(N)}_n g^n+\sum_{k=1}^\infty \sum_{l=1}^{k-1}\sum_{m=0}^\infty \left(\frac{i}{\sqrt{\pi}N!}\frac{2^{3N}}{g^{N+\frac{1}{2}}}e^{-\frac{2}{15g}}\right)^k \,\hskip -.1 cm  \left(\ln \left[-\frac{8}{g}\right]\right)^l \hskip -.1 cm c_{k, l, m}g^{m} \;,
\end{equation}
where the first summation term on the right hand side denotes the usual perturbative expansion ($\ref{pertcubic}$) of the $N$th excited state.

In \cite{Gahramanov:2015yxk} the functions $E(B,g)$ and $A(B,g)$ used in the quantization condition ($\ref{cubicquant}$) have been shown to satisfy the Dunne-\"{U}nsal relation in the following form (see also for the same relation \cite{Alv1})
\begin{equation}\label{DUrelation}
\frac{\partial E(B, g)}{\partial B} =  -\frac{g}{S} \left(B+ g \frac{\partial A(B, g)}{\partial g} \right) \;,
\end{equation}
where $E(B, g)$ is given by \cite{Gahramanov:2015yxk, Jentschura:2010zza}
\begin{align}\label{cubicenergy} \nonumber
E(B,g) & = B-g \left(\frac{7}{16}+\frac{15}{4} B^2\right)-g^2 \left(\frac{1155}{64}B+\frac{705}{16} B^3\right)\\ \nonumber
& \quad - g^3 \left(\frac{101479}{2048}+\frac{209055}{256} B^2+\frac{115755 }{128} B^4 \right) \\ 
& \quad - g^4 \left(\frac{129443349}{16384} B+\frac{77300685 }{2048} B^3+\frac{23968161}{1024} B^5\right) + \ldots \;.
\end{align}
One can then obtain the usual perturbation theory ground state energy ($\ref{pertcubic}$) by setting $B=\frac{1}{2}$ in the above expression.

Now let us calculate the fluctuation factors. For this reason we need to introduce the instanton fluctuation factor $H$ is given by (see e.g. \cite{ZinnJustin:1989mi})
\begin{align}\nonumber
H & = \exp{ \left[ -A+\frac{S}{g} \right]} \\ \label{fluctuationfactor} & = \exp{ \left[ S\int\frac{dg}{g^2}\left(\frac{\partial E}{\partial B}+\frac{Bg}{S}-1\right) \right]} \;,
\end{align}
where $S$ is the coefficient of the instanton action. Note that the second line in (\ref{fluctuationfactor}) is obtained by using the Dunne-\"{U}nsal relation (\ref{DUrelation}). Then we calculate the non-trivial fluctuation factor by the following equation \cite{Dunne:2013ada}
\begin{equation}
F = H \; \frac{\partial E(B, g)}{\partial B} \;.
\end{equation}
In fact, the coefficients of the function $F$ are the coefficients $c_{k, l, m}$ appearing in (\ref{cubicexpansion}) for $k=1, l=0$. 

By using the arguments above, we calculate the non-trivial fluctuation factor of one instanton sector for arbitrary $N$th energy level as
\begin{align} \nonumber 
F & \ = \ 1 + \left(-\frac{141 B^2}{8}-\frac{15 B}{2}-\frac{77}{32}\right) g\\
& \qquad  \ + \ \frac{\left(318096 B^4-223168 B^3-183864 B^2-186032 B-31031\right) }{2048}g^2+\ldots \;.
\end{align}
The imaginary part of the energy then can be written by
\begin{equation}
\text{Im} E_N = - \frac{ 2^{3N} e^{-\frac{2}{15g}}}{\sqrt{\pi} N! g^{N+\frac{1}{2}}}  F \;.
\end{equation}
Then by using the expansion ($\ref{cubicenergy}$), one can easily obtain the imaginary part of the energies for $N=0,1,2,3$ ($B=\frac12, \frac32, \frac52,\frac72$, respectively) as follows
\begin{align}
\text{Im} \; E_0 & = -\frac{e^{-\frac{2}{15g}}}{\sqrt{\pi g}}\left[1-\frac{169}{16}g-\frac{44507}{512}g^2-\frac{86071851}{40960}g^3-\frac{189244716209}{2621440}g^4-\ldots \right] \;, \\
\text{Im} \; E_1 & = -\frac{8e^{-\frac{2}{15g}}}{\sqrt{\pi g^{3}}} \left[ 1-\frac{853}{16}g+\frac{33349}{512}g^2-\frac{395368511}{40960}g^3-\frac{1788829864593}{2621440}g^4-\ldots \right] \;, \\
\text{Im} \; E_2 & = -\frac{32e^{-\frac{2}{15g}}}{\sqrt{\pi g^{5}}} \left[ 1-\frac{2101}{16}g+\frac{1823341}{512}g^2-\frac{1085785671}{40960}g^3-\frac{4272925639361}{2621440}g^4-\ldots \right] \;, \\
\text{Im} \; E_3 & = -\frac{256e^{-\frac{2}{15g}}}{3\sqrt{\pi g^{7}}} \left[ 1-\frac{3913}{16}g+\frac{8807869}{512}g^2-\frac{15716668611}{40960}g^3-\frac{3214761534593}{2621440}g^4-\ldots \right] \;.
\end{align}
They all agree with the results obtained in \cite{Kleinert:1995ii,JZcubic} (see also \cite{Alvarez1}) and can easily be extended beyond. Note that the imaginary part has a negative sign due to we are considering here resonance energy levels, whereas antiresonance energy levels have a positive imaginary part. Here one can justify the power of the formula (\ref{DUrelation}) For any state $N$, the perturbative expansion is enough to calculate the non-trivial fluctuation factor in the non-perturbative sector and contribution to the energy thereof.  

\subsection{Quartic anharmonic oscillator}
Similar results can also be obtained for the quartic anharmonic potential whose Hamiltonian we consider is 
\begin{equation}
\mathbb{H}(g) = -\frac{1}{2} \frac{\partial^2}{\partial q^2}  + \frac{1}{2}  q^2 +  g q^4.
\end{equation}

Note that for this potential the system has resonances on the real line of $g<0$ and instantons exists in this region. The decay width of a particle trapped on the local minima is not zero and complex part of the energy is associated with the instanton configurations. A single instanton action for this system reads $-\frac{1}{3g}$ which is positive. 

The Rayleigh-Schr\"{o}dinger perturbation theory yields the following energy expansion for the ground state energy \cite{Bender:1969si, Drummond}
\begin{equation}
E_{\rm ground} = \frac12 + \frac{3}{4} g - \frac{21}{8} g^2 +
\frac{333}{16} g^3 - \frac{30885}{128} g^4 + \ldots \;.
\end{equation}

Again the coefficients in this expansion grow like a factorial of the order in large orders and non-alternating by sign, which makes the series non-Borel summable\footnote{To our knowledge the application of summation methods to quartic anharmonic oscillator first was studied in \cite{Reid}. In \cite{Loeffel:1970fe} Loeffel et al. proved rigorously that the energy of quartic anharmonic oscillator with $g>0$ converge to the correct eigenvalues via Pade approximants. In \cite{Graffi:1990pe} Graffi et al. used Borel summation for the quartic potential with $g>0$ and proved the convergence of the result.}. 

The $N$th level excited state has the following resurgent expansion for the quartic potential
\begin{equation} \label{quarticexpansion}
E^{(N)}(g)=\sum_{n=0}^\infty E^{(N)}_n g^n+\sum_{k=1}^\infty \sum_{l=1}^{k-1}\sum_{m=0}^\infty \left(\frac{i}{\sqrt{\pi}N!}\frac{2^{2N+\frac{1}{2}}}{(-g)^{N+\frac{1}{2}}}e^{\frac{1}{3g}}\right)^k \,\hskip -.1 cm  \left(\ln \left[\frac{4}{g}\right]\right)^l \hskip -.1 cm c_{k, l, m}g^{m} \;.
\end{equation}
Again for this system, the perturbative and the non-perturbative functions satisfy 
\begin{equation}
\frac{\partial E(B, g)}{\partial B} =  -\frac{g}{S} \left(B+ g \frac{\partial A(B, g)}{\partial g} \right) \;,
\end{equation}
where $E(B, g)$ is given by  \cite{Gahramanov:2015yxk, Jentschura:2010zza}
\begin{eqnarray} \nonumber
E(B,g) \ & = & \ B +g \left(\frac{3}{8}+\frac{3 }{2} B^2 \right)  -g^2 \left(\frac{67 }{16} B+\frac{17}{4}B^3 \right) \\ \nonumber
& \quad & + g^3 \left(\frac{1539}{256}+\frac{1707}{32} B^2 +\frac{375}{16} B^4 \right)  \\ \label{quarE}
& \quad & - g^4 \left(\frac{305141}{1024} B+\frac{89165}{128} B^3+ \frac{10689}{64} B^5 \right) + \ldots \;.
\end{eqnarray}
The  usual perturbation series of the ground state can be regained by setting $B = \frac12$ in the above expression.

In case of quartic potential, the non-trivial fluctuation factor of one instanton sector for any $N$ is
\begin{align} \nonumber 
F & \ = \ 1+ \left(\frac{17 B^2}{4}+3 B+\frac{67}{48}\right) g\\
& \qquad \ + \ \frac{\left(41616 B^4-13248 B^3-31416 B^2-62640 B-14807\right) }{4608} g^2+\ldots \;,
\end{align}
which gives us the following imaginary part of the energies
\begin{align}
\text{Im} \; E_0 & = - \sqrt{-\frac{2}{\pi g}}e^{\frac{1}{3g}} \left[ 1+\frac{95}{24}g-\frac{13259}{1152}g^2+\frac{8956043}{82944}g^3+\ldots \right] \;, \\
\text{Im} \; E_1 & = - \sqrt{\frac{2^{5}}{\pi (-g)^{3}}}e^{\frac{1}{3g}} \left[1+\frac{371}{24}g  -\frac{3371}{1152}g^2+\frac{33467903}{82944}g^3+\ldots \right] \;,\\
\text{Im} \; E_2 & = - \sqrt{\frac{2^{9}}{\pi (-g)^{5}}} e^{\frac{1}{3g}} \left[ 1 + \frac{851}{24}g + \frac{262717}{1152}g^2 +\frac{69337223}{82944}g^3 + \ldots \right] \;,
\end{align}

for $N=0,1,2$ respectively. Imaginary part of the energies for $N=0,1$ have been computed in \cite{Jentschura:2010zza} by using both of the functions $A(E,g)$ and $B(E,g)$. Our expansions are in agreement with their results.

\section{Conclusions}
Unlike the first impression, already from the first coefficients of usual perturbative expansion one can get more information about an observable. Like non-perturbative effects to the system and large order correction terms to the perturbative expansion. The only question is how to extract the information out of it in a systematic way.

Here we considered cubic and quartic anharmonic potentials both of which have a local minima. Since the potentials are unbounded, a particle initially at the local minima will decay and its energy levels are complex. The complex part of the energy is associated with the instanton effects. Here we explicitly calculated the fluctuation terms of one-instanton sector for various energy levels by using the Dunne-\"{U}nsal relation and, therefore, we were able to write the imaginary part of energy for one-instanton contribution. Moreover, the instanton fluctuation terms we obtained here are related to the correction terms to the factorially growing large order coefficients of the perturbative sector by dispersion relations \cite{Jentschura:2010zza, Bender:1998ry}. Therefore already from the early terms of the perturbative expansion around a minima we can get information about the correction terms to the factorially growing large order coefficients of the same expansion.

Here we only derived one-instanton contributions to the energy eigenvalue but higher instanton contributions can be also calculated by using the perturbative data and the equation ($\ref{DUrelation}$) for both cubic and quartic anharmonic potentials. It would be interesting to extend this analysis to higher order anharmonic oscillators. So far it has only been shown that the  Dunne-\"{U}nsal relation works for potentials of order $\leq 4$. It has been pointed out in \cite{Gahramanov:2015yxk} that the Dunne-\"{U}nsal relation is not satisfied for higher order anharmonic potentials for given generalized quantization conditions. This opens up a possibility for a generalization of the relation. 
\vspace{0.4cm}

\textbf{Acknowledgments}\\
The paper is based on a talk given by KT at the International Conference on Quantum Science and Applications (ICQSA-2016),  May 25-27, 2016, Eskisehir, Turkey. KT thanks to  Gerald Dunne for many valuable discussions and suggestions. This work has been mostly completed at Max Planck Institute for Gravitational Physics (Albert Einstein Institute) and KT is very grateful for the hospitality of the Institute during his visit. IG would like to thank the Institut des Hautes Études Scientifiques, IHES (Bures-sur-Yvette, France) and Istanbul Center for Mathematical Sciences (Istanbul, Turkey) where a part of the work was done for the warm hospitality.

\end{document}